\documentclass[onecolumn,journal]{IEEEtran}
\pdfoutput=1
\IEEEoverridecommandlockouts
\usepackage{cite}
\usepackage{amsmath,amssymb}
\usepackage{amsfonts,bm}
\usepackage{algorithmic}
\usepackage{graphicx}
\usepackage{textcomp}
\usepackage{pgfplots} 
\pgfplotsset{compat=newest}
\usepackage{tikz}
\usepackage{tikzscale}
\usepackage{amsmath}

\usepackage{xcolor}
\def\BibTeX{{\rm B\kern-.05em{\sc i\kern-.025em b}\kern-.08em
    T\kern-.1667em\lower.7ex\hbox{E}\kern-.125emX}}

\newtheorem{lemma}{Lemma}

\newtheorem{example}{Example}
\newtheorem{remark}{Remark}
\newcommand{\wt}{ \mathrm{\, wt \,} }
\begin{document}

\title{On Decoding Using Codewords of the Dual Code}

\author{Martin Bossert, Fellow IEEE\\
\textit{Institute of Communications Engineering} \\
\textit{Ulm University}, Germany\\
martin.bossert@uni-ulm.de}

\maketitle

\begin{abstract}
We present novel decoding schemes  
for hard and soft decision decoding of block codes using the minimal weight
codewords of the dual code. 
The decoding schemes will be described for 
cyclic codes where polynomials 
can be used, however, the modification for non-cyclic codes is possible and straight forward. 
The hard decision decoding calculates syndrome polynomials
which are the product of the received polynomial with dual codewords.
Proper cyclic shifts of these syndrome
polynomials are obtained and 
the non-zero positions are counted componentwise for these shifts. 
The values of this counting are a reliability measure and can be used for
locating the error and also the non-error positions.
This reliability measure is the basis for various variants of hard decision decoding algorithms.
Decoding schemes with iterative error reduction
are possible as well as information set decoding using the inherent reliability information
of the measure even if there is no reliability information from the channel. 
Further,  we will show how reliability information from the channel can be included in order to obtain 
soft decision decoding schemes.
We derive the relation between bit flipping, believe propagation, and majority logic decoding 
to the novel schemes. 
As examples to illustrate the functioning 
we use BCH and Reed-Muller codes as examples for binary codes, and RS codes for non-binary codes.
We recall a known result that Reed-Muller codes punctured by one position are cyclic
and thus, are equivalent to special cases of BCH codes.
Simulation results for hard and soft decision decoding will be given for several examples
and compared with results from literature.
We show how Reed-Muller codes can be constructed and decoded using the
Plotkin construction. 
Finally, we analyze the soft decision decoding of the Plotkin construction 
and derive that one of the two codes uses a $3$ dB better channel (also known as channel polarization).

\end{abstract}

{\it Keywords:} Hard and soft decision decoding, iterative decoding, decoding using codewords of the dual code, decoding beyond half the minimum distance, cyclic codes, Plotkin construction, BCH codes, Reed-Muller codes, Reed-Solomon codes

\section{Introduction}
Non-algebraic
hard (HaDe) and soft decision (SoDe) decoding of block codes has a long history and numerous results exist.
In 1962 Gallager \cite{Gallager} introduced low-density parity-check (LDPC) codes
with bit flipping (BiFl) and believe propagation (BePr) decoding using extrinsic information.
After the rediscovery of LDPC codes at the end of the 90s the construction of LDPC codes and their SoDe decoding was subject of research for two decades.
Another basic approach was 1963 by Massey \cite{Massey} the threshold decoding
which is based on majority logic (MaLo) arguments where checks 
with a specific property are used. Namely, at
 the considered code position all checks must have a one
while at the other positions only one of the checks can have a one.
In 1986 a method was introduced in \cite{ferdi} 
using the minimal weight codewords of the dual code instead of only using the rows of the parity check matrix
in LDPC or the strong restriction in MaLo. Further, a combinatorial argument was given which is
supporting the functioning of HaDe and SoDe decoding.
Recently, in \cite{haeger} it was observed for Reed-Muller (RM) codes that 
using the minimal weight codewords with BePr gives improved decoding performance
compared to using only the rows of the parity check matrix.

There exist also SoDe methods in combination with a HaDe decoder and three of them are fundamental:
Forneys generalized minimum distance decoding (GMD)  \cite{Forney}, in 1966,  Chase decoding \cite{Chase}, in 1972,  
and Dorsch decoding\cite{Dorsch}, in 1974. 
According the reliability of a symbol GMD declares successively unreliable positions
as erasure and uses error erasure decoding for these.
The Chase algorithm modifies unreliable positions and each modification is 
decoded by a HaDe decoder.
Dorsch described that from the reliable received positions an information set can
be selected
and this can be encoded to a codeword and compared with the received vector.
Since the number of errors in the reliable positions
is small, modifications of the information set can be used to calculate
several candidates and choose the best.
The methods by Chase and Dorsch are also known under the name ordered statistic decoding.
In \cite{haeger} the method by Dorsch gives the best decoding performance. 

Another important result for decoding was by Plotkin \cite{Plotkin}, in 1960.  Even he gave a code construction later it was discovered in \cite{Schnabl}, in 1995 that the class of binary RM codes
can be HaDe and SoDe  decoded very efficiently using this construction.
Performance results of this decoding were also published in \cite{Lucas}. 
In \cite{Dumer} the idea of modifying the mapping for the Plotkin construction was presented.
These ideas were extended by Stolte \cite{Stolte} in 2002 where he introduced optimized 
codes for bitwise multistage decoding. In fact his constructions are very similar to Polar codes
by Arikan \cite{Arikan} in 2009. However, since Arikan showed that these codes are capacity achieving the interest for them grew rapidly. El-Khamy, et al. give in \cite{Khamy}  
simulation results for polar codes and describe their history. 
In \cite{Vardy} a concatenation of polar codes with a block code is introduced 
in order to select the decoding decision from a list. This construction has very good decoding performance.

Cyclic codes have the advantage that only a generator polynomial is necessary for encoding.
Here we will present a novel decoding concept for cyclic codes
based on the minimal weight codewords of the dual code. 
Syndrome polynomials are calculated by multiplying the received polynomial
by the dual codewords and  proper cyclic shifts of these syndromes are used for
calculating a reliability measure.
With this algorithm HaDe decoding beyond half the minimum distance is possible
as well as SoDe decoding.
Parts of these results has been published in \cite{SCC2019}. 

In Section \ref{basics} we give a brief introduction of the used code classes, namely
cyclic RS, BCH, and RM codes.
The Plotkin construction combines two codes in order to construct a code of double length.
The parameters of the resulting codes are derived and a decoder is given.
Then the class of RM codes is recursively constructed by the Plotkin construction.
We also recall an old result from 1968 (confirm \cite{Macwiliams}) that RM codes, punctured by one position, are 
equivalent to cyclic codes.
In fact these punctured RM codes can be interpreted as special cases of BCH codes, 
and thus, they can also be decoded by the presented method.
The basic idea of the decoding method is introduced in Section \ref{idea}.
We show how the method relates to BePr, BiFl, and MaLo. 
In additon, we analyze how the decoding benefits from the automorphisms of the used codes.  
Finally, we extend the combinatorial arguments from \cite{ferdi} and
calculate expected values and compare them with
simulated averages.
With this, no proof is possible that decoding beyond half the minimum distance 
works. However, it makes the functioning of the decoding plausible. 

In Section \ref{hard-dec} we describe possible HaDe decoding strategies
and apply them to binary and non-binary code examples.
The intension for these examples is not to have many simulation results 
for various applications but to show the possible potential of the presented decoding.  
First, we study the iterative error reduction for the BCH$(63,24,15)$ code
and compare simulated curves with bounded minimum distance decoding and 
interpolation based list decoding up to the Johnson radius.
Afterwards we compare iterative error reduction and information set decoding
for the punctured RM$(2,6)$ which is equuivalent to a BCH$(63,22,15)$ code.
We extend the decoding method to non-binary codes and as 
examples we consider the decoding  of a RS$(15, 5)$ and a  RS$(15, 11)$ code
over $\mathbb F_{2^4}$. Also here, decoding beyond half the minimum distance is possible.

Including reliability information from the AWGN channel is straight forward and introduced in 
Section \ref{soft-dec}. Simulation results are compared with those from \cite{haeger}.
Finally,  we consider SoDe decoding
when using the Plotkin construction with which codes from arbitrary classes can be combined.
The coderate of the resulting code is the average coderate of the used codes.
We proof that the code with larger rate views a channel which is 3 dB better 
than the original channel for the average coderate. 
This effect is also known as channel polarization.
We end with conclusions where we mention several open problems.

\section{Code Fundamentals}\label{basics}
For simplicity we restrict to binary extension fields $\mathbb F_{2^m}$  with a primitive element $\alpha$
(cf. \cite{Blahut} or \cite{Boss}).
Further, we only consider codes with primitive length $n=2^m-1$.
The  ring of polynomials is denoted by
$\mathbb F_{2^m} [x]$ and the quotient ring$\mod (x^n-1)$ by $\mathbb F_{2^m} [x]/(x^n-1)$. 
The polynomials are 
$c(x) = c_0 + c_1 x + \ldots + c_{n-1} x^{n-1}, c_i \in \mathbb F_{2^m}$.
Let the set $\{i_0, i_1, \ldots, i_{l} \}$ be the indices of the non-zero coefficients of $c(x)$. 
This set is called the support $\mathrm{supp} \, c(x)$.
The Hamming weight 
$\wt c(x)$ is the cardinality of its support,
  $|\mathrm{supp} \, c(x)| =\wt c(x)$.
  We will consider BCH, Reed-Muller (RM), and Reed-Solomon (RS) codes
  which all have a long history and possess many symmetries. 
  Since these code classes are well known we will introduce them shortly to give basic 
  notations and definitions.
  For further properties and proofs we refer to \cite{Blahut,Boss}, or \cite{Macwiliams}. 
\subsection{RS Codes}
Using  $\mathbb F_{2^m} $ and $n=2^m-1$ then a RS(n, k) code is e.g.\ obtained by choosing the powers 
$\alpha^{-j}, j = k, k+1, \ldots, n-1$
of the primitive element $\alpha$ as roots
of the generator polynomial. 
The generator polynomial $g(x)$ of this code is the product of linear factors
\begin{equation}\label{genpol}
g(x)= \prod\limits_{j=k}^{n-1}  (x - \alpha^{-j}).
\end{equation}
A possible way of encoding is the multiplication of information polynomials 
$i(x) = i_0 + i_1 x + \ldots + i_{k-1} x^{k-1}, i_j \in \mathbb F_{2^m}$
with the generator polynomial, $c(x)=i(x) g(x)$.
The minimum distance of the code is $d=n-k+1$ which fulfills the Singleton bound with equality and thus,
the code is maximum distance separable (MDS). So, given any $k$ positions 
of a codeword the remaining $n-k$ positions can be calculated.
The dual code is RS(n, n-k) with minimum distance $d^\perp=k+1$ and generator polynomial
$h(x) =\frac{x^n-1}{g(x)}$ which is also the parity check polynomial of the RS(n, k) code.

\subsection{BCH Codes}
We consider BCH codes as subfield-subcodes of RS codes. 
Let
the cyclotomic cosets be
	$K_i = \{i \cdot 2^j \mod n, j= 0,1, \ldots, m-1\}$
	where i is the smallest number in $K_i$.
Choosing $i=0,1,2, \ldots, n-1$ we get $n$ cosets, however,
two cosets are either identical or disjoint. The cardinality of the $K_i$ is $\leq m$. 
For any $K_i$ we can create an irreducible polynomial
\begin{equation}\label{monomials}
	m_i(x) = \prod\limits_{j \in K_i}  (x - \alpha^{-j})
\end{equation}
	which has coefficients from the base field $\mathbb F_2$. 
The roots of these polynomials are called conjugate roots since they create polynomials 
with coefficients from the base field $\mathbb F_2$. 
Now the generator polynomial is the product of different monomials $m_i(x)$.
The BCH$(n,k,d)$ code has length $n$, dimension $k=n - \mathrm{deg} g(x)$ and designed minimum distance is $d$
if $g(x)$ has $d-1$ consecutive roots.
Also here a possible way of encoding is the multiplication of information polynomials 
$i(x) = i_0 + i_1 x + \ldots + i_{k-1} x^{k-1}, i_j \in \mathbb F_{2}$
with the generator polynomial, $c(x)=i(x) g(x)$.
The dual code is a BCH$(n,n-k,d^\perp)$ with designed minimum distance $d^\perp$ and generator polynomial
$h(x) =\frac{x^n-1}{g(x)}$ which is also the parity check polynomial of the BCH$(n,k,d)$ code.

\subsection{RM Codes}
There exist various ways to define binary RM codes. We only describe two of them
which we need for the decoding schemes.
A RM code often is denoted by the order $r$ and $m$, RM(r,m) and is a binary code of length
$n=2^m$, dimension $k =\sum_{i=0}^r {m \choose i}$, and minimum distance $d=2^{m-r}$.

\subsubsection{Plotkin Construction of RM Codes}\label{Plotkin}
In 1960 Plotkin \cite{Plotkin}   
described a construction to create codes of length $2n$ using two codes of length $n$.
We will recall his construction for binary codes.
Let the codes $\mathcal C^{(1)}(n,k_1,d_1)$ and $\mathcal C^{(2)}(n,k_2,d_2)$ both $\subset \mathbb F^n_2$
be given. Then a code of doubled length $2n$,   dimension $k_1+k_2$, and minimum distance
$ \min\{ 2 d_1,  d_2\}$ can be constructed by
\[\mathcal C =\{ \mathbf{c} = (\mathbf{c}^{(1)} |\mathbf{c}^{(1)}+\mathbf{c}^{(2)}),\  \mathbf{c}^{(1)}\in \mathcal C^{(1)}, \mathbf{c}^{(2)}\in \mathcal C^{(2)}\}. \]
The length $2n$ and the dimension $k= k_1 + k_2$  are obvious.
For the minimum distance we describe a possible decoder of $\mathcal C$.
Assume a BSC  and $ \mathbf{r} =  \mathbf{c} +  \mathbf{e}$ is received. Due to the code construction
we can write
$ r= ( \mathbf{c}^{(1)} + \mathbf{e}^{(1)}| \mathbf{c}^{(1)}+ \mathbf{c}^{(2)}+ \mathbf{e}^{(2)})$ where $\mathbf{e}^{(1)}$ is the left half of $\mathbf{e}$ and $\mathbf{e}^{(2)}$  the right. 
In the first decoding step we add the left and the right half of $r$ and get 
$ \mathbf{c}^{(1)} +  \mathbf{e}^{(1)} +  \mathbf{c}^{(1)}+ \mathbf{c}^{(2)}+ \mathbf{e}^{(2)}
=  \mathbf{c}^{(2)} + \mathbf{e}^{(1)}+ \mathbf{e}^{(2)}$.
Since  $\mathrm{wt} (\mathbf{e}) 
\geq  \mathrm{wt} ( \mathbf{e}^{(1)} + \mathbf{e}^{(2)})$ we can decode  
$ \mathbf{c}^{(2)}$ correct if the number of errors $\tau =  \mathrm{wt} (\mathbf{e}) \leq \frac{d^{(1)}-1}{2}$.
In this case we know the correct $\mathbf{c}^{(2)}$ and can add this to the right half of $\mathbf{r}$ and we get
$ \mathbf{c}^{(1)} +  \mathbf{e}^{(2)}$ and  we have the left half $\mathbf{c}^{(1)} +  \mathbf{e}^{(1)}$.
Thus, we have the same codeword corrupted by two different errors, namely  $\mathbf{e}^{(1)}$ and $\mathbf{e}^{(2)}$.
According to the Dirichlet principle we can distribute $d^{(1)}-1$ errors in the two halves  
only such that either $ \mathbf{c}^{(1)} +  \mathbf{e}^{(1)}$ or $ \mathbf{c}^{(1)} +  \mathbf{e}^{(2)}$
contains $\leq \frac{d^{(1)}-1}{2}$ errors. If we decode both and select the one where we have corrected
the smaller number of errors we have the correct  $\mathbf{c}^{(1)}$ if the number of errors is $\leq d^{(1)}-1$.
This corresponds to a minimum distance of $2d^{(1)}$.
Thus,  the minimum distance of the constructed code is $ \min\{2 d_1,  d_2\}$.

If we start the Plotkin construction with the code $\mathcal C(2,1,2)$ and $\mathcal C(2,2,1)$
we can recursively
create the whole class of 
RM codes. With the first two codes we construct a 
$\mathcal C(4,3,2)$ code and we add the trivial codes $\mathcal C(4,1,4)$ and $\mathcal C(4,4,1)$.
With this we get the $\mathcal C(8,4,4)$ and the $\mathcal C(8,7,2)$ codes
and again can add the trivial codes $\mathcal C(8,1,8)$ and $\mathcal C(8,8,1)$ and so on.
In \cite{Boss} more details can be found and
in \cite{Schnabl}  recursive HaDe and SoDe decoding based on this construction was studied. 

\subsubsection{Cyclic Form of Punctered RM Codes}
In \cite{Macwiliams} Ch.13, \S 5, Th. 11 it is proven that 
the RM codes punctured by one position  are cyclic and thus are BCH codes with a particular choice
of cyclotomic cosets. Given the cyclotomic cosets $K_i$  for $n=2^m-1$.
The punctured RM(r,m) code is obtained when taking the monomials $m_i(x)$ according Eq. \ref{monomials}
as factors of the generator polynomial if the dual representation 
of $i>0$ has weight  $< m-r$.   
Thus, the RM code is a  BCH$(n,k,d=2^{m-r}-1)$ code.
The dual code is also a BCH$(n,n-k, d^\perp)$ with minimum distance $d^\perp$ and generator polynomial
$h(x) =\frac{x^n-1}{g(x)}$ which is also the parity check polynomial of the BCH$(n,k,d)$ code.

\begin{example}[R(2,6)]\label{punturedRM}
	Let $\alpha$ be a primitive element of $\mathbb F_{2^6}$. Let $n=2^6-1=63$.
Construct the cyclotomic cosets $K_i, i =0,1,3,5,7,9,11,13,15,21,23,27,31$.
The weight of the dual representation of the $i>0$ should be $< m-r=4$.
This is fulfilled for $K_i, i = 1=2^0,3=2^0+2^1,5= 2^0+2^2,7=2^0 +2^1+2^2,9=2^0+2^3,11
=2^0 + 2^1 +2^3,13=2^0+2^2+2^3,21=2^0+ 2^2+2^4$.
For $15 =2^0+2^1+2^2+2^3$ the weight is 4 and thus $K_{15}$ is not taken.
The product of corresponding irreducible polynomials gives the generator polynomial of degree
41. Thus $k=22$. Adding a parity check bit gives a codeword of the RM(2,6) code (64,22,16)   
The designed minimum distance of the BCH code is 15 and the parity bit gives 16.
\end{example}
Thus, all RM codes punctured by one position can be viewed as particular cyclic BCH codes
and only a generator polynomial is necessary for encoding.

\section{Decoding Approach}\label{idea}
We assume a binary symmetric channel (BSC)
with error probability $p$. A codeword  $c(x) \in \mathcal C(n,k,d)$ is transmitted
and $r(x)=c(x)+e(x)$ is received where $e(x)$ is the error polynomial. 
In order to decode, the dual code $\mathcal C^\perp(n,n-k,d^\perp)$ can be used.
The parity check matrix $H$ of a code is the generator matrix of the dual code. It consists of $n-k$
linearly independent codewords (or checks) from the dual code. Particular sets of checks are used for 
different decoding concepts, namely, BiFl, 
MaLo, and InSe decoding.
The soft decision variants, BePr, weighted MaLo, and InSe will be considered later.  
All these sets of checks consist of codewords of the dual code which can be created by linear combinations of the rows of $H$. 
In case of cyclic codes any non-zero  codeword of the dual code is divisible by the 
parity check polynomial $h(x)$ and
a polynomial multiplication of the received polynomial $r(x)=c(x)+e(x)$ with a 
codeword of the dual code$\mod (x^n-1)$ corresponds to $n$ checks.
A choice of any $n-k$ of these $n$ checks is a parity check matrix.
Here, we will use all cyclically different minimal weight codewords of the dual code as checks.
This choice was first used in \cite{ferdi}, however, here we present a novel interpretation.
Recently, in \cite{haeger} minimal weight codewords of the dual were used for decoding of RM codes. 
We start describing the basic idea of the decoding strategy and then 
show relations to BiFl and MaLo decoding.
Several properties of the decoding strategy will be given and analyzed.
Finally, we will give a plausibility analysis for the functioning of this 
decoding using combinatorial arguments.  
This analysis is not a proof 
that decoding beyond half the minimum distance is possible
but it strongly supports this property.

\subsection{Basic Idea}
If not stated otherwise all polynomials in this section are elements of $\mathbb F_{2^m} [x]/(x^n-1)$. 
Let $b(x) = x^{b_0} + x^{b_1} + \ldots + x^{b_{d^\perp-1}}$ be a dual codeword of weight $d^\perp$.
Since the code is cyclic we can assume $b_0=0$, thus, the coefficient at $x^0$ is $1$.
The support of this polynomial is  $\mathrm{supp} (b(x))=\{0, b_1, \ldots, b_{d^\perp -1}\}$
where $b_j \in \{1,2, \ldots, n-1\}$.
According the definition of the dual code $c(x) b(x) =0 \mod (x^n-1)$  holds for all codewords of $\mathcal C$.
Let the error be $e(x)=x^{e_0} +  x^{e_1} + \ldots + x^{e_{\tau-1}}$ 
and the received polynomial is $r(x)=c(x)+e(x)$.
The polynomial $w(x)$ is the product of the dual codeword $b(x)$ with the received polynomial
$r(x)$. In fact, the $w(x)$ can be considered as syndrome since it only depends on the error. 
Clearly, $w(x)$ is identical to the product of $b(x)$ 
with the error $e(x)$, since 
\begin{equation}\label{bconvolution}
w(x)= r(x) b(x) = (c(x) + e(x))b(x) = c(x) b(x) +  e(x)b(x) 
= e(x) b(x) \mod (x^n-1).
\end{equation}
A possible interpretation of $w(x)$ is the addition of cyclic shifts of the error $e (x)$ (where coefficients at the same position $x^i$ are added in $\mathbb F_2$)
\[
	\begin{array}{rcl}
		w(x) &=& x^{b_0} e(x) +  \ldots + x^{b_{d^\perp-1}} e(x) \mod (x^n-1)\\
		     &=& x^{e_0} +  x^{e_1} + \ldots + x^{e_{\tau-1}} +\\
		     & & x^{e_0+ b_1} +  x^{e_1 + b_1} + \ldots + x^{e_{\tau-1} +b_2} +\\
   & \vdots & \\
		 & & x^{e_0+ b_{d^\perp-1}} +  x^{e_1 + b_{d^\perp-1}} + \ldots + x^{e_{\tau-1} +b_{d^\perp-1}}, 
	\end{array}
\]
where the exponents $ e_i + b_j$ are calculated$\mod n$.
Note, that  for all
$e (x) \not\in \mathcal C$ the polynomial $w(x) \in \mathcal C^\perp$ is a
non-zero codeword of the dual code and therefore, $d^\perp \leq \wt w(x) \leq \min\{\tau d^\perp ,n\}$.
Any non-zero coefficient of $w(x)$ is an error (at its original position) or a shifted error.
We can shift $w(x)$ by the values $b_j \in \{-b_1, -b_2, \ldots, -b_{d^\perp-1}\}$ 
and, including $w(x)$, we have $d^\perp$ polynomials $x^{b_j} w(x) \mod (x^n-1)$.  
Since shifting does not change the weight, any non-zero coefficient of $w(x)$ is 
at the original error position in one of these shifts.
 In other words, in the set of all $d^\perp$ shifts we have at least $\wt w(x)$ errors at their original position.
This fact is obvious since all shifted $b(x)$ have the form $1+ \ldots$.
That we have at least $\wt w(x)$ errors at their original position is due to the fact that a shifted error (which stays non-zero in $w(x)$) can be eventually also at
another error position in some shift.  
Note, that the following relation  holds  
\[
	x^{b_j} w(x) = x^{b_j} (b(x) e(x)) = (x^{b_j} b(x)) e(x) \mod (x^n-1)
\]
thus, we only need to shift $w(x)$ and not multiply by the shifted $b(x)$.
We use the BCH(63,24,15) as an example to illustrate these 
properties. This code is still the best code known with this parameters.
\begin{example}[Shifts of $w(x)$ for BCH(63,24,15)]
	A codeword of the dual code
with minimum weight $d^\perp=8$ is
	$b(x)= x^{49} + x^{37} + x^{34} + x^{30} + x^{19} + x^{12} + x^6 +1$.
Assume the error is $e(x)= x^{42} + x^{38} + x^{11}$.
Then the polynomial $w(x) = e(x) b(x) \mod (x^n-1)$ is
$w(x) = x^{61} +x^{60} + x^{57} +x^{54} +x^{50} + x^{45} +x^{44} + x^{42} + x^{41} + x^{38} + x^{30} + x^{28} +x^{24} + x^{23} + x^{17} + x^{16} + x^{13} +x^{12} + x^{11} + x^{5}.  $
The positions  $x^{9}$ and  $x^{48}$ have disappeared but all other 
non-zero positions are error or shifted error positions.
For example the shift $x^{-19} w(x)$ contains all three error positions $x^{61-19}= x^{42}$,
$x^{57-19}= x^{38}$, and $x^{30-19}= x^{11}$.
\end{example}

\smallskip
The main idea is now, to count the number of ones in each position $j$
of all $d^\perp$ shifts of $w(x)$ and we denote this number by $\Phi_j$.
Shifting $w(x)$ by $x^{-b_i}$ means that the coefficient at $j+b_i  \mod n$ is at position 
$j$ after the shift. Therefore, we get
\begin{equation}
{ \Phi_j} =  \sum\limits_{i \in \mathrm{supp} \, b(x)} \  w_{j+i\mod n},\ j=0, 1, \ldots, n-1. 
\end{equation}
Recall, that the coefficient $w_j$ is the addition in $\mathbb F_2$ of coefficients of $r(x)$, namely 

\begin{equation}\label{polcheck}
w_j= r_j + r_{j-b_1} + r_{j-b_2} + \ldots + r_{j-b_{d^\perp -1}}. 
\end{equation}
Instead of only a single codeword we can use $L$ minimum weight codewords $b^{(\ell)}(x), \ell=0, \ldots, L-1$ of the dual code
which are cyclically different. Codewords are cyclically different if no shift $i$ exists such that 
$x^i b^{(\ell_1)}(x) = b^{(\ell_2)}(x) \mod (x^n-1), \ell_1 \neq \ell_2$.
Then, for position $j=0, 1, \ldots, n-1$  the counting becomes
\begin{equation}\label{shiftsum}
{ \Phi_j} = \sum\limits_{\ell=0}^{L-1} \sum\limits_{i \in \mathrm{supp} \, b^{(\ell)}(x)} \  w^{(\ell)}_{j+i\mod n}, 
\ , 
\end{equation}
where the value is bounded by $ 0 \leq \Phi_j \leq L d^\perp$.
With (\ref{polcheck}) and (\ref{shiftsum}), the polynomial
multiplication, the shifting, and the counting for position $j=0, 1, \ldots, n-1$ can be done by
\begin{equation}\label{phi_coeff}
	\Phi_j= \sum\limits_{\ell=0}^{L-1} \sum\limits_{i=0}^{d^\perp-1}  \left(\sum\limits_{l=0}^{d^\perp-1}
	r_{(j+ b^{(\ell)}_i - b^{(\ell)}_l\mod n)}\mod 2 \right).
\end{equation}
Recall, that  $b^{(\ell)}_0=0$.
It is convenient to introduce the sets 
of parity check supports
$\mathcal P (j,\ell, i)$ 
defined by
\begin{equation}\label{pcpos}
	\mathcal P (j,\ell,i) =\{ j + b^{(\ell)}_i -  b^{(\ell)}_l\mod n, \ l= 0,1, \ldots, d^\perp-1\}.
\end{equation}
Note, that the position $j$ is included in each  
$\mathcal P (j,\ell, i)$ 
since  $b^{(\ell)}_i -  b^{(\ell)}_l =0$ for $l=i$ 
independent of $\ell$. 
Thus, we have  $Ld^\perp$ check equations for each position $j$ with the property
that position $j$ is included in each of these check
equations and the remaining $d^\perp-1$ check positions are from the $n-1$ other positions.
A possible decoding strategy is to calculate $\Phi_j$ for each position $j$.
Then find the maximal value $\Phi_{j_m}$  and flip the position $j_m$ 
 by adding $x^{j_m}b^{(\ell)}(x)$ to $w^{(\ell)}(x)$.
If this addition results in zero an error is found.
This decoding strategy will work if the error positions have larger values  $\Phi_j$ than non-error 
positions. The following example illustrates this effect.
\begin{example}[Simulation of $\Phi$ Values]
We use the $L=35$ cyclically different codewords
of minimum weight $d^\perp=8$ of the dual of the BCH(63,24,15) code. 
The simulation uses $2000$ random errors of weight $\tau$
and calculates $\Phi$.
For $\tau=5$, in all 2000 cases  all the error positions correspond to the $5$ largest values of $\Phi$
and for $\tau=6$ in 1999 to the $6$  largest values.
In $1884$ cases the largest $7$ values correspond to the error positions for $\tau=7$.
For $\tau=8$ in 830 cases the largest $8$ values of $\Phi$ correspond to the error positions.
Note, that $\tau=8$ is beyond half the minimum distance.
\end{example}

\subsection{Relations to Known Decoding Strategies}
In order to compare vector based descriptions to polynomial based descriptions, we first derive the
connection between them. 
Given the dual codeword $b(x)=1 + x^{b_1} + \ldots + x^{b_{d^\perp-1}}$ the vector  $\mathbf p_t$ with support $\{n-1,n-1-b_1, \ldots, n-1-b_{d^\perp-1}\}$ is a codeword of the dual code. This reverse operation
is necessary to relate the polynomial multiplication to the scalar product (compare Eq. \ref{phi_coeff}). 
The scalar product of $\mathbf p_t$ with the received vector $\mathbf r$ is a parity check  
$\langle \mathbf p_t, \mathbf r \rangle =  r_{n-1}+ r_{n-1-b_1}+ \ldots
+ r_{n-1-b_{d^\perp-1}}$
and results in $0$ or $1$.
Thus, each coefficient $w_i$ of $w(x)=r(x) b(x)$ is the result of the scalar product of the corresponding 
dual codeword $\mathbf p_t$ with $\mathbf r$, 
for example $w_{n-1}= r_{n-1}+ r_{n-1-b_1}+ \ldots + r_{n-1-b_{d^{\perp-1}}}$.
By cyclically shifting the  $L$ codewords we can create $n L$
different minimum weight codewords and thus,  $\mathbf p_t, t =0, \ldots, nL-1$ parity checks. 
Clearly, the $nL$ coefficients of $w^{(\ell)}(x), \ell =0, \ldots, L-1$ are the results of the 
parity checks $\langle \mathbf p_t, \mathbf r \rangle, t =0, \ldots, nL-1$.

\subsubsection{Relation to MaLo}
In MaLo, decoding of position $j$, a set of parity checks ($\mathbf p_i, i=1, \ldots, J$) is used which 
has the following property \cite{Boss}:
at position $j$ all checks are $1$ and at
all other positions  only one of the checks has a $1$. 
In other words $\cap_i \mathrm{supp}(\mathbf p_i )=j$.
If the majority of the checks is 
not fulfilled ($=1$), position $j$ is decided to be erroneous. 
Note that for $\mathrm{wt} (\mathbf p_i)= d^\perp$ it follows that the 
number $J$ of checks which can exist is bounded by $J\leq \lfloor (n-1)/(d^\perp-1) \rfloor$.
Thus, the $\Phi_j$ calculated according to Eq. \ref{pcpos} can be interpreted as
modified MaLo decoding. Position $j$ is $1$ in each of the $d^\perp L$ checks. However, the other positions are 
$1$ in more than one check, since $d^\perp L> \lfloor (n-1)/(d^\perp-1) \rfloor$. 
Because of this  violation, no proof that a certain number of  errors can be corrected is possible
as for MaLo decoding.
However, to use more checks for the voting about position $j$ turns out as an advantage.
So, if the majority of the checks for $j$ are not fulfilled, which is counted by $\Phi_j$,
the position is considered as erroneous. 

\subsubsection{Relation to BePr}
Fixing $j,\ell$, and $i$ in Eq. \ref{pcpos} 
we get the support of a check equation, which we denote by $\mathcal P(j, \ell,i)=\{j, t_1, t_2, \ldots t_{d^\perp-1}\}$.
The check equation is  $r_j + r_{t_1} + \ldots +r_{t_{d^\perp-1}}=s_j \in \{0,1\}$.
The extrinsic information $\hat r_j$ for position $j$ is $\hat r_j = r_{t_1} + \ldots +r_{t_{d^\perp-1}} =s_j+r_j$ and is the information of the other positions about position $j$.
Clearly, when $s_j=0$ it follows that the extrinsic information is $\hat r_j = r_j$ and for  
$s_j=1$ it is $\hat r_j \neq r_j$. For decoding with the extrinsic information we count 
 the number of $s_j=1$ when $\ell=0,\ldots,L-1$ and $i=0,\ldots,d^\perp-1$ which is $\Phi_j$.
Thus, the presented decoding can be interpreted as using the extrinsic information 
of $d^\perp L$ check equations for a position $j$  
and therefore, is identical to MaLo decoding.

\subsubsection{Relation to BiFl }
In BiFl position $j$ in the received vector $\mathbf r$ is flipped resulting in the vector 
$\mathbf r_j$ and 
 the scalar products $\langle \mathbf p_t, \mathbf r \rangle$  are compared
 with $\langle \mathbf p_t, \mathbf r_j \rangle$. If the number of scalar products with result
 $1$ is reduced, position $j$ is considered as erroneous.
This can be described using the $w^{(\ell)}(x)$ as follows.
Flipping position $j$ is the addition of $x^j b^{(\ell)}(x) \mod (x^n-1)$ to $w^{(\ell)}(x)$ 
and counting the number of different positions. Note, that this is equivalent to the correlation
between $x^j b^{(\ell)}(x) \mod (x^n-1)$ and $w^{(\ell)}(x)$. 
In order to measure the change in the scalar products for BiFl we define the value $\Delta_j$ by
\begin{equation}\label{delta}
\Delta_j = \sum\limits_{\ell=1}^L \mathrm{wt} (w^{(\ell)}(x) + x^j b^{(\ell)}(x)) - \mathrm{wt} (w^{(\ell)}(x)).
\end{equation}
\begin{lemma}[Relation between $\Phi$ and $\Delta$]\label{bfmalo}
The relation between the presented decoding strategy and bit flipping is $L d^\perp - 2\Phi_j= \Delta_j$.
\end{lemma}
\begin{IEEEproof}
	We use only one polynomial $w(x) = e(x) b(x)\mod (x^n-1)$.
The value  of $\Delta_j$ is only dependent on the non-zero positions of 
$x^j b(x)$ which are $\mathcal J =\{j, j+b_1, j+ b_2, \ldots, j+ b_{d^\perp-1}\}$. 
If the value of $w_i,\  i \in \mathcal J$ is $1$ then $-1$ is added to $\Delta_j$ and if the value is $0$
then $1$ is added. The $\Phi_j$ is calculated by shifting $w(x)$
by $0,-b_1, \ldots, -b_{d^\perp-1}$, which is $\Phi_j= \sum_{j \in \mathcal J} w_j$ 
and corresponds to the number of ones in these $d^\perp$ positions.
The number of zeros is then $d^\perp -\Phi_j$ and 
$\Delta_j$ is the number of zeros minus the number of ones 
thus, $\Delta_j =  (d^\perp -\Phi_j) - \Phi_j $.
This is valid for each of the $L$ polynomials, which completes the proof.
\end{IEEEproof}
The decoding strategies MaLo and BiFl using minimal weight codewords from the 
dual code will make the same decoding decision since the smallest  $\Delta_j$ corresponds 
to the largest $\Phi_j$ which can be interpreted as the extrinsic information for position $j$.
The calculation of $\Phi_j$ uses $L d^\perp$ checks therefore,  $n(L d^\perp)$ checks
are used for all positions. There are $n L$ checks for BiFl, however, for each check
$d^\perp$ extrinsic information are calculated which results in  $d^\perp(n L)$ checks.

\subsection{Useful Property}
A bijective linear mapping $f: \mathcal C \rightarrow \mathcal C $ of a code onto itself is called automorphism. 
The group of all automorphisms is 
denoted by $\mathrm{Aut}(\mathcal C)$.
Two examples of automorphisms for cyclic codes are cyclic shifting and squaring$\mod (x^n-1)$. 
If we shift all codewords
by $j$ positions, we again get all codewords and shifting does not change the weight. 
The same holds for squaring.
Also squaring of codewords with coefficients from $\mathbb F_{2^m}$
is weight preserving,
$\mathrm{wt}(c(x))=\mathrm{wt}(c^2(x)) \mod (x^n-1)$ (cf. \cite{Macwiliams,Boss}).
A consequence is, that for all $f \in \mathrm{Aut}(\mathcal C)$ and for all $c(x) \in \mathcal C$ 
we have $f(c(x)) \in \mathcal C$.
We can apply any automorphism of the code to errors and can calculate further
errors of the same weight. The number of errors we can create depends on the number of automorphisms
of the used code.
\begin{lemma}[Number of Automorphisms for binary BCH Codes]
The number of automorphisms of binary BCH codes of length $n=2^m-1$ is at least  $mn$.
\end{lemma}
\begin{IEEEproof}
	There might be additional automorphisms besides shifting and squaring.
	Using shifting we can calculate $n$ errors (including the unshifted error).
	Each of them we can square $m$ times (including the power $1$). 
\end{IEEEproof}
	Given an error $e(x)$ of weight  $\tau$ then  $mn$ errors of the same weight can be calculated
	by applying shifting and squaring.
	Thus, we can calculate $nm-1$ additional errors from an error $e(x)$.
However, it can happen that $f(e(x)) =e(x)$, which means that the set of different errors   
is less than $nm$.
\begin{remark}The squaring can be used when searching for minimal weight codewords. Especially, when
considering binary BCH codes 
of length $n=2^m-1$. If we have found one $b(x)$ of weight $d^\perp$, then  
$(b(x))^2$, $(b(x))^{2^2}$, $\ldots$, $(b(x))^{2^{m-1}}$ are almost all cyclically 
different other minimal weight codewords (all calculations$\mod (x^n-1)$). 
Recall, that $(b(x))^{2^{m}} = b(x)$.
\end{remark}

Given a $\Phi$ we denote by   $\mathrm{sort}(\Phi)$ the vector which is sorted 
according the values, $\Phi_i$.
We say two different $\Phi_a$ and $\Phi_b$ have the same coefficient distribution
if $\mathrm{sort}(\Phi_a) = \mathrm{sort}(\Phi_b)$. 
With these facts we can state the following Lemma.
\begin{lemma}[Property of $\Phi$]
The $\Phi$ according to Eq. \ref{phi_coeff}, which is calculated from an error $e(x)$, has the same coefficient distribution as the $\Phi$ calculated by $f(e(x))$ for $f \in \mathrm{Aut}(\mathcal C)$.
\end{lemma}
\begin{IEEEproof}
	Consider the set $\mathcal B$ of all minimal weight codewords $\mathbf p_j, j=0, \ldots, nL-1$, according 
	to the last subsection. 
	For any automorphism we have $f(\mathcal B)=\mathcal B$.
Therefore, it holds that
\[
	\Delta_s(\mathbf e)=\sum\limits_j \langle \mathbf p_j, \mathbf e \rangle = 
\sum\limits_j \langle f(\mathbf p_j), f(\mathbf e) \rangle =
\sum\limits_j \langle \mathbf p_j, f(\mathbf e) \rangle. 
\]
The same holds for  $\Delta_s(\mathbf e_j)$ where we have flipped one bit in position $j$.
The $\Delta_j$ from
Eq. \ref{delta} are the difference between $\Delta_s(\mathbf e)$ and $\Delta_s(\mathbf e_j)$.
Lemma \ref{bfmalo} relates bit flipping to $\Phi$. 
So the values are the same, however, the positions are permuted. Thus, it is the same coefficient distribution.
\end{IEEEproof}
For our decoding strategy, this means that if we can correct an error $e(x)$,
we most likely can correct the error $f(e(x))$.
This statement might be wrong in cases where the maximal values in $\Phi$ are not unique.
In other words, for errors of weight $\tau$ not ${n \choose \tau}$  
different coefficient distributions for $\Phi$ exist, but a smaller number $({n \choose \tau})/(mn)$. 
This fact supports the combinatorial arguments considered next.
\subsection{Plausibility Analysis}\label{plausible}
First we calculate the expected value of the weight $\omega$ of $w(x)$ as in \cite{ferdi}. 
Let $\mathbf p_t$ be a parity check vector and $\mathbf e$ the error vector.
The result of the scalar product $\langle \mathbf p_t, \mathbf r \rangle =\langle \mathbf p_t, \mathbf e \rangle$ depends on the particular choice of the check $\mathbf p_t$ 
and the error $\mathbf e$ and is not known. However, we can calculate 
the number of scalar products with result $1$ of one check $\mathbf p_t$ with all possible errors
of weight $\tau$. Since the scalar product is $1$, if an odd number of errors intersect
with the $d^\perp$ ones in  $\mathbf p_t$, we can count all these cases.
We define the number $W \in \mathbb N$ dependent on $ \tau, d^\perp$, and $n$ by
\begin{equation}
W=  { d^\perp \choose 1} {n- d^\perp \choose \tau- 1} +{ d^\perp \choose 3} {n- d^\perp \choose \tau- 3} + \ldots +q,
\end{equation}
where 
\[
q= \left\{ \begin{array}{ll}
{ d^\perp \choose \tau} {n- d^\perp \choose 0}, & \tau \ \mathrm{odd}\\
{ d^\perp \choose \tau-1} {n- d^\perp \choose 1}, & \tau \ \mathrm{even}.
\end{array}
\right.
\]
Thus, the expected weigth of one error of weight $\tau$ with one  $\mathbf p_t$ is
$W/ {n \choose \tau}$. Because $w(x)$ corresponds to $n$ checks, the expected
weight $\omega$ is
\begin{equation}\label{DA}
E[\omega] =  \frac{n W}{{n \choose \tau}}.
\end{equation}
Therefore, in each shift of $w(x)$ we have in average at least $E[\omega]/d^\perp$ errors
at the original position of the $\tau$ errors, while the remaining $E[\omega]-E[\omega]/d^\perp$
positions are at the $n-\tau$ non-error positions.
The summation of the $d^\perp$ shifts of $w(x)$ 
calculates the frequency of occurrence of ones at the $n$ code positions 
where the summation of the coefficients is done as integer addition. 
Clearly, the value $\Phi_j$ is the number of ones at position $j$ in the $d^\perp$
shifts of $w(x)$.
Since shifting does not change the weight, 
the expected number of errors in
any of the $ d^\perp$ shifts  is $E[\omega]/d^\perp$.
Thus, the probability that an error position is $1$ is $E[\omega]/(\tau d^\perp)$.
The expected number of ones in $w(x)$ is then
 $E[\omega]/\tau $.
At an error position $j$,
the frequency of occurrence $\Phi_j$
 has then the expected  value
\begin{equation}\label{phierror}
	E[\Phi_e(\tau)] =  \frac{E[\omega]}{\tau} L.
\end{equation}
For a non-error position, the probability for a $1$ is $d^\perp (E[\omega] - E[\omega]/d^\perp)/(n - \tau)$
and the expected value of  $\Phi_j$
at a non-error position is
\begin{equation}\label{phicorrect}
	 E[\Phi_c(\tau)] =   \frac{d^\perp(E[\omega] - \frac{E[\omega]}{d^\perp})}{n-\tau}   L.
\end{equation}
In the following example these expected values are compared with simulated averages.
\begin{example}[Comparison of  Expected Values and Simulated Averages]
Again we use the $L=35$ different codewords
of minimum weight $d^\perp=8$. The simulation uses $2000$ random errors of weight $\tau$
and calculates the parameters in Table \ref{tabelle}.
The expected weight $E[\omega]$ is calculated by Eq. \ref{DA}
and simulated as $AV[\omega] $ and the values are almost identical.
The value $AV[\Phi_e(\tau)]$ is the average of an error position
in the simulation
and $E[\Phi_e(\tau)]$ is predicted by Eq. \ref{phierror}. For a non-error position the value is predicted
by Eq. \ref{phicorrect}.
It can be observed that the estimated values for non-error positions are better than those for the
error positions. This is due to the above mentioned effect that a $w_j=1$, 
which corresponds to a shifted error at one position can, in addition, be 
shifted to other error position.

\begin{table}[h]\caption{\label{tabelle}Predicted and measured parameters of $\Phi$}
\begin{center}
\begin{tabular}{|c|c|c|c|c|c|}
$\tau$ & $5$& $6$ & $7$ & $8$ &
	$9$   \\
\hline 
\hline
$E[\omega] $ & $25.2 $& $27.2 $& $28.6 $& $29.6 $& $30.3 $\\
$AV[\omega] $ & $25.2 $& $27.2 $& $28.6 $& $29.6 $& $30.3 $\\[1ex]
\hline 
$E[\Phi_e(\tau)]$ & $181.5 $& $163.0 $& $146.9 $& $133.3 $& $121.3 $\\
$AV[\Phi_e(\tau)]$ & $192.2 $& $179.5 $& $169.5 $& $162.3 $& $156.7  $\\[1ex]
\hline 
$E[\Phi_c(\tau)]$ & $109.5 $& $118.4 $& $128.6 $& $135.7 $& $153.0 $\\
$AV[\Phi_c(\tau)]$& $108.6 $& $120.1 $& $125.7 $& $131.3 $& $135.6 $\\[1ex]
\hline 
 $E[\Phi_{max}]$ & $196.8 $& $188.9 $& $181.6 $& $175.9 $& $141.5 $\\
\hline

\end{tabular}
\end{center}
\end{table}
\end{example}

\section{Hard Decision Decoding}\label{hard-dec}
According Section \ref{plausible}, the values $\Phi_j$ can be used as reliability measure
for position $j$, even in case of a channel with no reliability information.
Large values indicate an error and small ones a non-error position.
Therefore, we will describe two decoders which both use this reliability information.
The first one will reduce the number of errors iteratively by flipping the $\mu$ positions with the largest 
values $\Phi_j$ and
the second will use the $k+k_0$ positions with the smallest values $\Phi_j$ to find an information set
of size $k$. The value $k_0$ has to be chosen such that the corresponding $k \times (k+k_0) $ submatrix of the
generator matrix has rank $k$.

Both variants need to calculate the $\Phi_j$ according to Eq. \ref{phi_coeff}.
When storing the $n L d^\perp$ supports $\mathcal P (j,\ell,i)$, the complexity reduces to
$n L (d^\perp)^2$ XOR operations and $n L d^\perp$ integer increments. 
Now we must find the $\mu$ largest values of $\Phi$, which has a complexity of
$\mu n$ comparisons or $(k+k_0) n$ in case of information set decoding, which needs the smallest values.

1) Iterative error reduction: 
Find the $\mu$ positions $j_1, j_2, \ldots, j_\mu$ in $\Phi$ which have the largest values 
$\Phi_{j_1} \geq \Phi_{j_2} \geq \Phi_{j_3} \geq \ldots \geq \Phi_{j_\mu}$. These are probably the error positions.
Now these positions are flipped by calculating for $\ell=0, \ldots, L-1$ the new polynomials $\hat w^{(\ell)}(x) = w^{(\ell)}(x) + x^j b^{(\ell)}(x)$ successively for $j_1, j_2, \ldots, j_\mu$.
Note, if during the calculation $\hat w^{(\ell)}(x)=0$ an error which corrects to a valid codeword
was found.

2) Information set decoding:
Find the $\nu=k+k_0$ positions $j_1, j_2, \ldots, j_\nu$ in $\Phi$ which have the smallest values 
$\Phi_{j_1} \leq \Phi_{j_2} \leq \Phi_{j_3} \leq \ldots \leq \Phi_{j_\nu}$. These are probably the non-error positions.
This requires $\nu \cdot n$ comparisons.
Now the smallest $k$ positions are used as systematic positions and if the rank of the 
matrix consisting of the columns of the generator matrix has full rank this is possible. 
If the rank is smaller columns are exchanged by the $k_0$ extra positions.
Then a list of $k$ candidates are calculated by encoding with the systematic positions 
where each of these bits are flipped.

\subsection{Binary Codes} 
\begin{figure}
 \centering
\includegraphics[width=0.4\textwidth]{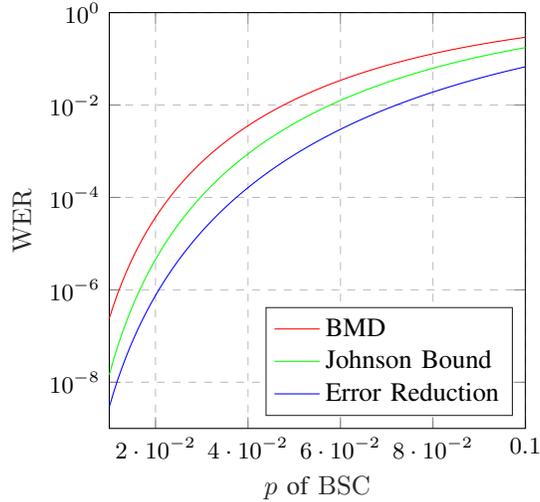}
\caption{\label{hard63}WER for BCH(63,24,15) versus  error 
probability $p$ of a BSC} 
\end{figure}

In the following we will give simulation results for two examples, namely the BCH$(63,24,15)$ code
and the punctured RM$(2,6)$ code described in Example \ref{punturedRM}.
In the first example we use iterative error reduction and compare it to the
best known algebraic list decoding. The second example  compares information set decoding
to iterative error reduction. 
Further examples for the decoding of BCH codes using all minimal weight codewords of the dual code 
for BiFl can be found in \cite{ferdi} and for Quadratic Residue codes in \cite{QRcodes}. 
\begin{example}[Bitflipping for the BCH$(63,24,15)$ code]
For the dual BCH code we have $L=35$ cyclically different codewords of weight $d^\perp=8$. 
We use $\mu=7$ and simulate random errors of weight $\tau=5, 6, \ldots, 15$
and calculate the probability that an error can be corrected. With these probabilities and those
that $\tau$ errors occur in a BSC with error probability $p$ the word error rates (WER) 
can be calculated and is shown in Fig. \ref{hard63}. For comparison the WER for bounded minimum 
distance decoding  and the binary Johnson bound  are also plotted.
The bound from \cite{venkatscript} is the number of errors ($\tau=8$) which can be corrected by 
algebraic interpolation based list decoding. Note, that for list decoding it was assumed that
the decision for $8$ errors is always correct.
At $p=0.05$ the new decoding algorithm has WER$=0.00083$ while BMD WER$=0.013$
and list decoding WER$=0.004$.
\end{example}
Possible improvements for the new decoding algorithm are: i) run several decoders with different values of $\mu$,
which gives in general a list of decoding decisions. ii) Use the idea of Chase algorithm  
by flipping all of the $\mu$ positions for $\mu$ decoders and all ${\mu \choose 2}$ pairs  
and run a decoder for each, and so on.
iii) Use the reliability information for believe propagation or the selection of particular checks.  

\begin{figure}
 \centering
\includegraphics[width=0.4\textwidth]{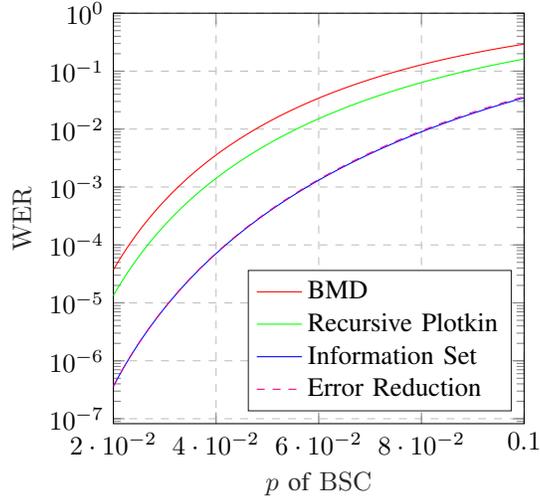}
\caption{WER for Punctured RM(63,22,15) versus  error 
probability $p$ of a BSC \label{RM63}} 
\end{figure}

\begin{example}[Cyclic punctured RM Code]
	We use a BCH$(63,22,15)$ code which is equivalent 
	the punctured  RM$(2,6)$  (see Example \ref{punturedRM}). There are $L=155$ minimal weight 
	$d^\perp=8$ dual codewords.
	The strategy for error reduction is here to flipp all positions with the maximal value $\Phi_j$
	which can be one or more positions, so $\mu$ is adaptive.
	For information set decoding a proper set of $k$ positions are chosen
	out of the smallest  values $\Phi_j$. Then
		list of $k+1$ candidates is calculated by encoding with the these $k$ positions
		unchanged and flipped.
From the list the one with the smallest Hamming distance is taken and if the decision is not
unique a random choice is done.
	In Fig. \ref{RM63}, the information set decoding is compared to iterative error reduction.
Both methods have the same performance and so only one curve is visible.
	For comparison the BMD decoding, which can correct up to $7$ errors, and the 
recursive decoding of the  RM$(2,6)$ using the Plotkin construction,  are also plotted.
Note, that the BCH$(63,22,15)$ has a slightly better performance than the BCH$(63,24,15)$
which is due to the fact that the number of minimal weight dual codewords is smaller for the latter one. 
\end{example}

\subsection{Non-binary Codes}
For non-binary codes we also use the cyclically different minimal weight dual codewords of the form 
$b(x)=1 + \beta_1 x^{b_1} + \ldots+ \beta_{d^\perp -1} x^{d^\perp -1}$,
where $\beta_i \in \mathbb F_{2^m}$.
The errors in an q-ary symmetric channel are non-binary and therefore the error is
$e(x)=\varepsilon_1 x^{e_1} + \ldots +\varepsilon_\tau x^{e_\tau}$.
Here, $w(x)= e(x) b(x) \mod (x^n-1)$  has terms of the form $\beta_i\varepsilon_l x^{e_l+b_i}$. 
If we shift this term by $(\beta_i)^{-1} x^{-b_i}$ we get  $\varepsilon_l x^{e_l}$
and have not only the error position but also the error value.
Therefore, we can use different counters for each value $0, \alpha^i, i=0,1, \ldots, n-1$,
which results in a matrix $\mathbf \Phi = \Phi_{i,j}$, where $i$ represents the field element  
$\alpha^i, i=0,1, \ldots, n-1$. The value for $w_j=0$ we count in  $\Phi_{n,j}$.
The columns $j$ represent the positions $0,1, \ldots n-1$.
Suppose we have $L$ cyclic different $b^{(\ell)}(x)$, then it holds that
$\sum_i \Phi_{i,j} = L d^\perp$. In other words $\Phi_{n,j} = L d^\perp - \sum_{i \neq n} \Phi_{i,j} $.
Note that a value equal to zero is more likely a  non-error position than an error position.

\begin{figure}
 \centering
\includegraphics[width=0.4\textwidth]{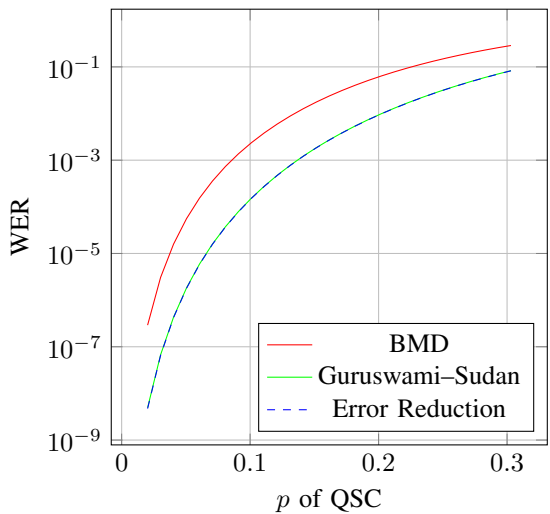}
\caption{WER for RS$(15, 5)$ versus  error 
probability $p$ of a q-ary symmetric channel \label{RS15}} 
\end{figure}

Several decoding strategies are possible.
The first is to find the maximum $\Phi_{i_{m_l},j}$ of each column of $\mathbf \Phi$ excluding the last row.
Then find the largest values $\Phi_{i_{m_1},j_1} \geq \Phi_{i_{m_2},j_2}\geq \ldots$ and subtract the error
$\alpha^{i_{m_l}}$. With this we get the error positions and the error values.
Another strategy is to find the minimal values $\Phi_{n,j_l}$ of the last row  of the matrix $\mathbf \Phi$,
which are more likely error positions. Also a combination of both strategies is possible.
We will first give an example describing the resulting values
and a second example with simulations.
\begin{example}[RS$(15, 5)$ Code]
We decode the RS$(15, 5)$ code by using the $335$ cyclic different minimal weight $d^\perp=6$  
codewords of the dual  RS$(15, 10)$ code.
The Johnson radius for this code is $\tau_J = n-1- \lfloor \sqrt{n(k-1)}\rfloor =7$
which means that interpolation based list decoding can correct two errors more than 
half the minimum distance.
The new decoding for a random example with $\tau=5$ errors gives the following values.
The maximal  $\Phi_{i_{m_l},j}$ values of each column are 
$(136, 142, 134,$ $140, 140, 137,$ $ 138,  317,  325,$ 
$312, 131, 313,$ $309, 151, 131)$
one can see the significant difference of values larger than $300$ which are error positions
and values less than $152$ which are non-error positions.
The error positions are $(7,8,9,11,12)$.
The corresponding error values are $\alpha^3, \alpha^0, \alpha^{11}, \alpha^2, \alpha^8$.
The last row of matrix $\mathbf \Phi$, which counts the number of zero values, is
$(216, 209, 204, 209, 206, 211, 194, 107, 113, 120, 206, 103, 113, 201, 210)$,
which has the five smallest values $\leq 120$ at the error positions and the 
non-error position have values $\geq 194$.
For $\tau=6$ these numbers are for the maximal value of each column for 
error positions $\geq 203$ and for non-error positions $\leq 169$. For the last row 
the values for error positions are $\leq 130$ and those for non-error positions $\geq 143$.
For a random example with $\tau=7$ we observe that the four largest maxima are error
positions, thus, the error is correctable.
And even more, for $\tau=8$ we observed  examples in which the largest values are error
positions, thus, the error is correctable.
In Fig. \ref{RS15}, the simulation results are plotted. The WER curve for the new
decoding algorithm is identical
to using the Guruswami-Sudan list decoding algorithm up to the Johnson radius. 
The implementation of the later was from \cite{sage}
 and when the list contained several solutions with the same number of errors a random choice
was done. For comparison the WER values for BMD decoding are included. 
\end{example}

\begin{example}[RS$(15, 11)$ Code] 
	The RS$(15, 11)$ code can be decoded by the $31$ cyclically different
	dual codewords of weight $12$. Here the Johson bound is $\tau_J=2$,
	which is half the minimum distance and the performance of interpolation based list decoding
	is the same as for BMD.
	With the presented decoding method all errors of weight $\leq 2$ and $3 \%$ of 
	the errors of weight $\tau=3$ can be corrected. 
\end{example}

The class of non-binary BCH codes can also be decoded with the presented decoding method. 
The advantages compared to RS codes are, that the minimal weight of the dual code is smaller,
since this codes are not MDS, which improves the decoding performance 
and the number of minimal weight codewords is smaller, which reduces complexity.

\section{Soft Decision Decoding}\label{soft-dec}
We assume BPSK modulation $x_j \in \{1,-1\}$ with the mapping of 
code to modulation symbols $0 \leftrightarrow 1$ and $ 1 \leftrightarrow -1$.
The Gaussian channel adds noise $n_j \in \mathcal N (0, \sigma^2)$ with zero mean and variance $\sigma^2$.
We receive $\mathbf y = \mathbf x + \mathbf n$.
The hard decision of $\mathbf y$ is $\mathbf r$ with $r_j = 0$ if $y_j > 0$
and $r_j = 1$ if $y_j < 0$.

\subsection{Including Reliability Information}
According to Eq. \ref{phi_coeff}, for each position $j$ we get $Ld^\perp$ parity check equations, which
all  contain the position $y_j$. 
For a fixed $j, \ell$ and $i$ according  to Eq. \ref{pcpos} we get the support   
$\{ j, t_1, t_2  \ldots, t_{d^\perp-1}\}$.
For soft decision decoding we use
extrinsic information for position $j$ derived from the values
$y_{t_1}, y_{t_2}, \ldots,  y_{t_{d^\perp-1}}$.
The sign of the product of these positions $\mathrm{sign}(y_{t_1} \ldots  y_{t_{d^\perp-1}})$
is the same as the sign of $y_j$ if $\mathrm{sign}(y_j y_{t_1} \ldots  y_{t_{d^\perp-1}})=1$
and different if $\mathrm{sign}(y_j y_{t_1} \ldots  y_{t_{d^\perp-1}})=-1$.
Considering only the sign is hard decision decoding.
Therefore, concerning the sign, there is no difference
between weighted MaLo and BePr decoding.
We need to calculate a reliability information for the check
given the values $y_j,y_{t_1}, y_{t_2}, \ldots,  y_{t_{d^\perp-1}}$.
There exist several methods for the calculation of this reliability information and 
the choice influences the decoding performance.
In this work we only want to show the principle of 
including soft information into the presented decoding algorithm.
Therefore, we choose the reliability information 

\begin{equation}\label{extrinsic}
	\rho(j,\ell,i) = \frac{1}{|y_j|}\mathrm{sign} \left( y_{t_1} \ldots  y_{t_{d^\perp-1}}\right)
	\mathrm{min} \{  |y_{t_1}|, \ldots,  |y_{t_{d^\perp-1}}|\}.
\end{equation}
The extrinsic information 
$\varphi_j$ for a position $j$ for all checks is then
\begin{equation}\label{majo}
	\varphi_j = \sum\limits_{\ell=0}^{L-1} \sum\limits_{i=0}^{d^\perp-1} \rho(j,\ell,i),
\end{equation}
where the values $\varphi_j < 0$ are likely errors and the values $\varphi_j > 0$ are likely non-errors.
A simple algorithm is to calculate $\varphi_j$ and to flip the $\mu$ positions with negative 
and largest absolute values and stop if the syndrome is zero.
We do not update the received values $y_j$, but only change the sign to flip the bit.
It was observed that  the largest positive values are a good indicator for non-error positions which can be used for information set decoding.

\begin{figure}
	\begin{center}
	\includegraphics[width=0.5\textwidth]{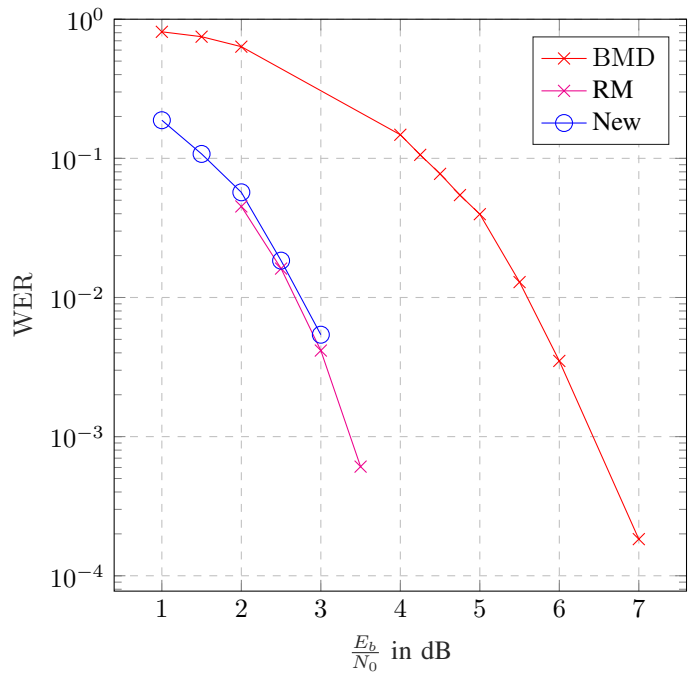} 
	\caption{WER for soft decision decoding of the BCH$(63,24,15)$ and RM$(3,7)$ code\label{softcurve}}
	\end{center}
\end{figure}

\begin{example}[Soft Decoding of the BCH$(63, 24, 15)$ Code]
For the dual BCH code we have $L=35$ cyclically different codewords of weight $d^\perp=8$. 
In Fig. \ref{softcurve}, the simulation results for $\mu=7$ and reliability information according to Eq. 
\ref{extrinsic} is shown. At a WER of $6\cdot 10^{-3}$, 
a gain of about 3 dB compared to hard decision BMD is achieved.
For comparison the results of decoding a RM code of length $128$
with minimal weight codewords from Fig. 1 of \cite{haeger} is plotted.
It can be seen that the results are very similar, even the code for the presented 
decoding has only half the length.
In addition, improvements seem possible when using list decoding and/or when using an advanced 
calculation to obtain the reliability information.  
Further,  Chase-like variants and information set decoding are possible.
\end{example}

\subsection{Plotkin Construction and Soft Decision Decoding}\label{plotkin}
The novel decoding method allows soft decision decoding, which can be used in the Plotkin construction 
\cite{Plotkin}. With this construction
a code can be created based on two other codes having a lower and a higher code rate.
The created code has the average code rate of the two component codes used
which determines the signal to noise ratio of the channel.
In the following we will derive that these component codes are virtually transmitted over different channels. 
We proof that one code has a $3$ dB better signal to noise ratio than the channel for the average code rate.
We transform the construction from Section \ref{Plotkin} into BPSK, which gives 
$\mathbf x  = (\mathbf x^{(1)} |\mathbf x^{(1)} \odot \mathbf x^{(2)})$.

Transmitting this codeword over an AWGN channel,  
we receive $\mathbf y = ( \mathbf y^{(1)}| \mathbf y^{(2)}) = (\mathbf x^{(1)} + \mathbf n^{(1)}
|\mathbf x^{(1)} \odot \mathbf x^{(2)} + \mathbf n^{(2)})$, where the noise is normal distributed with  zero mean and variance $\sigma^2$, therefore $n_i^{(1)},n_i^{(2)} \in \mathcal N (0, \sigma^2)$. 

\subsubsection{Decision for $\mathbf x^{(2)}$}
In the first step we need to get an estimate of the symbols $x_i^{(2)}$ of the codeword $\mathbf x^{(2)}$.  The approximation used in \cite{Schnabl} is
$ \mathrm{sign}(y_i^{(1)} y_i^{(2)} ) \mathrm{min} \{|y_i^{(1)}|, |y_i^{(2)}| \}$. 
With this reliabilities a decoder for  $\mathcal C^{(2)}$ gives  an estimate for $\mathbf x^{(2)}$.

\subsubsection{Decision for $\mathbf x^{(1)}$}
\begin{figure}
	\begin{center}
	\includegraphics[width=0.4\textwidth]{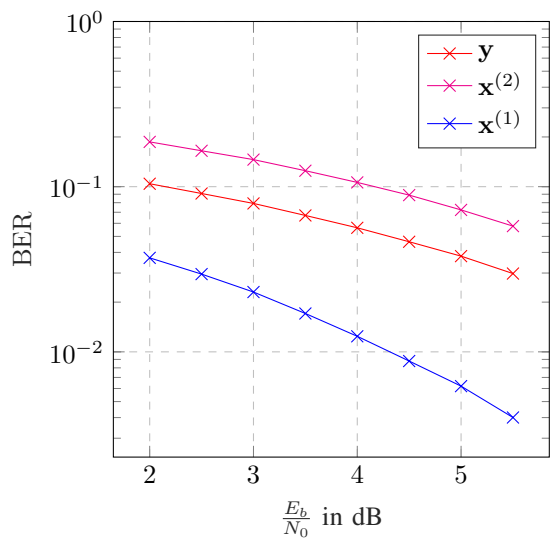}
\caption{Gaussian Channel BER for Plotkin Construction\label{3dbgain}}
	\end{center}
\end{figure}

Assume that  $\mathbf x^{(2)}$ was correct decoded, then we can multiply  $\mathbf y^{(2)} \odot \mathbf x^{(2)}$ and calculate
\begin{equation}\label{GMC}
\mathbf y^{(1)} + \mathbf y^{(2)} \odot \mathbf x^{(2)}=
\mathbf x^{(1)} + \mathbf x^{(1)} \odot \mathbf x^{(2)} \odot \mathbf x^{(2)} + \mathbf n^{(1)} + \mathbf n^{(2)} \odot \mathbf x^{(2)} = 2 \mathbf x^{(1)} + \mathbf n^{(1)} + \tilde{\mathbf{n}}^{(2)}. 
\end{equation}
Since the amplitude is doubled the signal power is four times increased, while the 
noise power is only doubled. Thus, a gain of factor two ($3$ dB) is obtained.
In Fig. \ref{3dbgain}, the error probability of the channel and for the two codes 
are visualized for an overall coderate of $R=1/2$. 
\begin{example}[Channels Using $4$ Codes]
	Using the Plotkin construction twice we get 
	$\mathbf x  = (\mathbf x^{(1)} |\mathbf x^{(1)} \odot \mathbf x^{(2)}|
	\mathbf x^{(1)} \odot \mathbf x^{(3)}|\mathbf x^{(1)} \odot \mathbf x^{(2)}\odot \mathbf x^{(3)}\odot \mathbf x^{(4)})$.
	For a coderate of $R=1/2$ for the overall code at an AWGN with $E_b/N_0 = 2$ dB,
the bit error probability of the channel is $p_c=0.104$.
The code $4$ has a channel with bit error probability  $p_4= 0.302$
while $p_3= 0.101$, $p_2= 0.072$, and  $p_1= 0.006$. The last code has a channel which is $6$ dB better.
Note that for the construction of RM codes $\mathbf x^{(2)}$ and $\mathbf x^{(3)}$
are different codewords of the same code.
\end{example}
\section{Conclusions}
We have presented a decoding strategy to decode beyond half the minimum distance, by using the minimal weight codewords from the dual code. 
The improvement, compared to BMD decoding,is considerable and was shown for a BCH code.
The calculated values contain inherent reliability 
information, which can be used for information set decoding even for channels 
without reliability information. 
This was shown
for a punctured RM code which is cyclic and equivalent to a special BCH code.
The decoding works also for non-binary codes and gives error positions as well as error values
which was demonstrated for RS codes of different rates.
Reliability information from the channel can be included in decoding which improves the decoding performance
and together with Plotkin construction splits a channel into a good and a bad one. 
Note that improvements of the decoding performance
of the presented ecoding method are possible when using list decoding
by running parallel decoders with different choices of the maximal/minimal values to start
for both, soft and hard decision decoding.
However, many open other questions also remain.
SoDe decoding of nonbinary BCH and RS codes, adaptive selection of the checks used for decoding, 
selection of codes used in SoDe of Plotkin constructions, 
dependence of the decoding performance on the selection of cyclotomic cosets for BCH codes,
decoding performance of the presented decoding with puncturing, only to mention a few.
\section*{Acknowledgement}
The author would like to thank Sebastian Bitzer, 
Rebekka Schulz, and Jiongyue Xing 
for helpful discussions and the simulation of 
several codes and  decoding schemes. Further thanks for discussions go to
Sven M\"uelich, Cornelia Ott, Sven Puchinger, and Carmen Sippel.

\end{document}